\documentclass[sigconf]{acmart}
\usepackage{graphicx}
\usepackage{textcomp}
\usepackage{xcolor}
\usepackage{booktabs}
\usepackage{colortbl}
\usepackage{color}
\usepackage{xcolor, colortbl}
\definecolor{grayone}{gray}{.9}
\definecolor{graytwo}{gray}{.7}
\usepackage{xspace}
\usepackage{url}
\usepackage{balance}
\usepackage{tabulary}
\usepackage{array}
\usepackage{graphicx}
\usepackage[ruled,lined,linesnumbered,vlined,algo2e]{algorithm2e}
\usepackage{subcaption}
\usepackage{listings}
\usepackage{lipsum}
\usepackage{hyperref}
\usepackage{graphicx}
\usepackage[T1]{fontenc}
\usepackage[ansinew]{inputenc}
\usepackage[many]{tcolorbox}
\usepackage{extarrows}
\usepackage{booktabs}
\usepackage{multirow}
\usepackage{colortbl}
\usepackage{tabularx}
\usepackage{enumitem}
\setitemize[0]{leftmargin=15pt}
\graphicspath{{fig/}}
\usepackage[justification=centering]{caption}
\usepackage{pifont}
\usepackage{makecell,multirow,diagbox,rotating}
\usepackage{longtable}
\usepackage{etoolbox} 
\usepackage{listings}
\usepackage{algorithm}
\usepackage{algorithmicx} 
\usepackage{algpseudocode}
\usepackage{amsmath}
\usepackage{dutchcal}
\usepackage{breakurl}
\usepackage{url}
\definecolor{codegreen}{rgb}{0,0.6,0}
\definecolor{codegray}{rgb}{0.5,0.5,0.5}
\definecolor{codepurple}{rgb}{0.58,0,0.82}
\definecolor{backcolour}{rgb}{0.95,0.95,0.92}
\makeatletter
 \def\@textbottom{\vskip \z@ \@plus 1pt}
 \let\@texttop\relax
\makeatother
\lstdefinestyle{mystyle}{
	backgroundcolor=\color{backcolour},   
	commentstyle=\color{codegreen},
	keywordstyle=\color{magenta},
	numberstyle=\tiny\color{codegray},
	stringstyle=\color{codepurple},
	basicstyle=\ttfamily\footnotesize,
	breakatwhitespace=false,         
	breaklines=true,                 
	captionpos=b,                    
	keepspaces=true,                 
	numbers=left,                    
	numbersep=5pt,                  
	showspaces=false,                
	showstringspaces=false,
	showtabs=false,                  
	tabsize=2
}

\usepackage{cleveref}
\usepackage[many]{tcolorbox}
\usepackage{pifont}

\usepackage{makecell}
\usepackage{threeparttable}
\makeatletter  
\newif\if@restonecol  
\renewcommand\footnoterule{%
	\kern-3\p@
	\hrule\@width\columnwidth
	\kern2.6\p@}

\usepackage{url}

\SetCommentSty{mycommfont}

\usepackage[font=normalsize,skip=3pt]{caption}
\newcommand{\distance}{3pt}
\definecolor{Green}{RGB}{0,180,0}
\newcommand{\ling}[1]{\textcolor{magenta}{#1}}

\newcommand{\chengwei}[1]{\textcolor{violet}{#1}}

\def\BibTeX{{\rm B\kern-.05em{\sc i\kern-.025em b}\kern-.08em
    T\kern-.1667em\lower.7ex\hbox{E}\kern-.125emX}}

\copyrightyear{2022}
\acmYear{2022}
\setcopyright{acmcopyright}\acmConference[ICSE '22]{44th International Conference on Software Engineering}{May 21--29, 2022}{Pittsburgh, PA, USA}
\acmBooktitle{44th International Conference on Software Engineering (ICSE '22), May 21--29, 2022, Pittsburgh, PA, USA}
\acmPrice{15.00}
\acmDOI{10.1145/3510003.3510142}
\acmISBN{978-1-4503-9221-1/22/05}

\begin{document}

\title{Demystifying the Vulnerability Propagation and Its Evolution via Dependency Trees in the NPM Ecosystem}

\author{Chengwei Liu}
\orcid{0000-0003-1175-2753}
\affiliation{%
  \institution{College of Intelligence and Computing, Tianjin University}
  \city{Tianjin}
  \country{China}
}
\additionalaffiliation{%
  \institution{Nanyang Technological University}
    \city{Singapore}
  \country{Singapore}
}
\email{chengwei001@e.ntu.edu.sg}

\author{Sen Chen}
\orcid{0000-0001-9477-4100}
\authornote{Sen Chen is the corresponding author.}
\affiliation{%
  \institution{College of Intelligence and Computing, Tianjin University}
  \city{Tianjin}
  \country{China}}
\email{senchen@tju.edu.cn}

\author{Lingling Fan}
\affiliation{%
  \institution{College of Cyber Science, Nankai University}
  \city{Tianjin}
  \country{China}
}
\email{linglingfan@nankai.edu.cn}

\author{Bihuan Chen}
\affiliation{%
 \institution{School of Computer Science and
Shanghai Key Laboratory of Data
Science, Fudan University}
 \city{Shanghai}
 \country{China}}

\author{Yang Liu}
\affiliation{%
  \institution{School of Computer Science and Engineering, Nanyang Technological University}
  \city{Singapore}
  \country{Singapore}}

\author{Xin Peng}
\affiliation{%
  \institution{School of Computer Science and
Shanghai Key Laboratory of Data
Science, Fudan University}
  \city{Shanghai}
  \country{China}}

\begin{abstract}
Third-party libraries with rich functionalities facilitate the fast development of JavaScript software, leading to the explosive growth of the NPM ecosystem. However, it also brings new security threats that vulnerabilities could be introduced through dependencies from third-party libraries. In particular, the threats could be excessively amplified by transitive dependencies. Existing research only considers direct dependencies or reasoning transitive dependencies based on reachability analysis, which neglects {the} NPM-specific dependency resolution rules as adapted during real installation, resulting in wrongly resolved dependencies. Consequently, further fine-grained analysis, such as precise vulnerability propagation and their evolution over time in dependencies, cannot be carried out precisely at a large scale, as well as deriving ecosystem-wide solutions for vulnerabilities in dependencies.

To fill this gap, we propose a knowledge graph-based dependency resolution, which resolves the inner dependency relations of dependencies as trees (i.e., \textsf{dependency trees}), and investigates the security threats from vulnerabilities in dependency trees at a large scale. Specifically, we first construct a complete dependency-vulnerability knowledge graph (\textsf{DVGraph}) that captures
the whole NPM ecosystem (over 10 million library versions and 60 million well-resolved dependency relations). Based on it, we propose a novel algorithm (\textsf{DTResolver}) to statically and precisely resolve dependency trees, as well as transitive vulnerability propagation paths, for each package by taking the official dependency resolution rules into account. Based on that, we carry out an ecosystem-wide empirical study on vulnerability propagation and its evolution in dependency trees. Our study unveils lots of useful findings, and we further discuss the lessons learned and solutions for different stakeholders to mitigate the vulnerability impact in NPM based on our findings. For example, we implement a dependency tree based vulnerability remediation method (\textsf{DTReme}) for NPM packages, and receive much better performance than the official tool (\textsf{npm audit fix}).
\end{abstract}



\maketitle

\section{Introduction}\label{sec:introduction}
Due to the rapid growth of functionality complexity in software applications, software componentization has become an irresistible trend in software development, leading to the boosting of third-party libraries.
As investigated, over 1.7 million Node.js libraries have been published on NPM \cite{npmjs.org, npmlibrariesio} (a node package manager) to facilitate software development.
As Contrast Security~\cite{contrastsecurity} revealed, third-party libraries appear in a majority (79\%) of today's software~\cite{williams2012unfortunate}.
However, every coin has two sides. Although using libraries reduces development cost and time, these integrated libraries pose a new security threat to the software ecosystem in practice, that vulnerabilities in these libraries may expose software 
{that depend on them}
 under security risks constantly~\cite{decan2018impact,zimmermann2019small,zhan2021atvhunter,zhan2020automated}. For example, \textsf{lodash}~\cite{lodash}, a widely-used JavaScript utility library with over 80 million downloads as dependencies per month, is identified to have severe vulnerability of prototype pollution and exposes 4.35 million projects on GitHub to the potential risk of being attacked~\cite{lodashimpact}.

Previous works have investigated vulnerability impact across the NPM ecosystem, while their approaches either only statically consider direct dependencies~\cite{decan2018impact},
or excessively analyze dependencies based on static reachability reasoning~\cite{zimmermann2019small} which may introduce inaccurate transitive dependencies ({illustrated by the motivating example of \Cref{fig:npmexample} (b) in \Cref{subsec:motivating}}) resulting in false-positive vulnerability warnings. 
None of the existing approaches provide precise dependencies, especially the inner complex relations among dependencies of software, at a large scale, which makes the impact of their analysis weakened and limits further solutions (i.e., precise remediation) to be proposed.
{Although some existing SCA tools (e.g., Snyk~\cite{snyk} and Blackduck~\cite{blackduck}) support NPM dependency analysis for user projects, most of them retrieve dependency trees from real installation rather than static reasoning.}
Besides, dependencies, as well as vulnerabilities in dependencies, are actually under dynamic change over time due to the flexibility of semantic versioning~\cite{semver}. 
Therefore, although existing work has also investigated the impact of vulnerabilities~\cite{decan2018impact,zimmermann2019small},
it is still challenging to analyze the evolution of vulnerability propagation existing in dependencies at a large scale without static and precise dependency resolution,
not to mention to derive practical solutions on preventing vulnerabilities from dynamically being introduced into dependencies.

To fill these gaps, we face the following challenges.
1) {{\textbf{Completeness}}}. 
NPM ecosystem is the largest platform with over 1 million published packages, which are hard to be fully analyzed.
{Some existing work only either analyzed with a limited number of libraries~\cite{wittern2016look, kula2017impact, dey2018software}, or studied vulnerabilities in dependency trees of limited projects~\cite{javanjafari2021dependency, chinthanet2021lags, alfadel2020threat}. }
Besides, NPM allows various ways to reuse third-party libraries as dependency constraints~\cite{npmpackagejson}, which are also hard to fully capture and resolve.
2) {{\textbf{Accuracy}}}. 
Existing work~\cite{wittern2016look,kula2017impact,decan2016topology,lertwittayatrai2017extracting} only conduct dependency-based analysis by identifying transitive dependencies via reachability reasoning while neglecting the NPM-specific dependency resolution rules~\cite{npmalgo}, which would lead to inaccurate results. 
3) {{\textbf{Efficiency}}}. 
Even though real installation can precisely retrieve dependency trees, installing NPM packages (i.e., \textsf{npm install}) is known to always take minutes per run~\cite{npminstalllong}, 
which is obviously not efficient enough to support large-scale studies. 
4) {\textbf{Dynamic updates}}. {NPM packages are known to have the most dependencies. 
{Any new release of libraries in dependency trees could lead to changes in installed dependencies for installation afterwards (as the example in \Cref{fig:vulintroexample}),}
which even complicates the management of dependencies, as well as the vulnerability propagation in dependencies.} Thus, it is challenging to explore the vulnerability propagation evolution over time. 

\begin{figure*}
	\centering
	\includegraphics[width=0.9\textwidth]{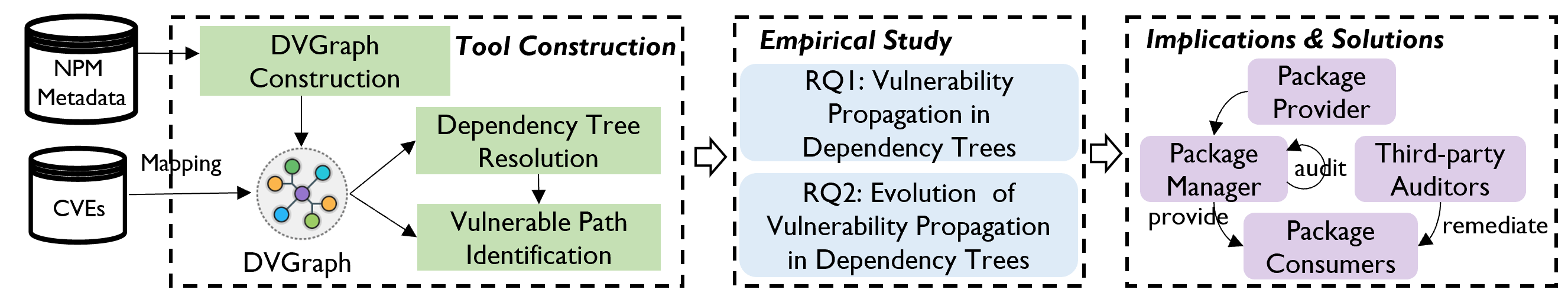} 
	\caption{Overview of our work}
	\label{fig:approach}  
\end{figure*}

In this paper, to overcome these challenges, 
\textbf{1)} we implement \textsf{a robust dependency constraint parser} to tackle the {diversity} of NPM dependency constraints, and based on it, we construct a {complete} dependency-vulnerability knowledge graph (\textsf{DVGraph}) to capture {the dependency relations among} all NPM packages {(over 1.14 million libraries and 10.94 million versions)}, as well as over 800 known CVEs (Common Vulnerabilities and Exposures)~\cite{cve} from NVD~\cite{nvd}, which further supports the thorough analysis of vulnerability propagation.
\textbf{2)} We propose an accurate {DVGraph based dependency} resolution algorithm (\textsf{DTResolver}) to calculate {dependency trees}\footnote{The resolved dependency graph when a given root package is installed.}
{at any installation time.} 
{It integrates the official dependency resolution rules and DVGraph-based reasoning to simulate the process of installation ({illustrated by the motivating example of \Cref{fig:npmexample} (c) in \Cref{subsec:motivating}}). 
Our \textsf{DTResolver} is validated to have an accuracy of over 90\% of resolved dependency trees being exactly the same comparing to real installation.}
(3) We further conduct an empirical study on vulnerability propagation in dependency trees. First, we investigate the characteristics of dependency trees brought by NPM dependency resolution ({details are available at our supplementary material}), based on which, 
we analyze the impact and features of vulnerability propagation in dependency trees, 
{particularly the vulnerabilities from transitive dependencies. Besides, we also extend our study to time dimension to investigate the evolution of vulnerability propagation in dependency trees over time to unveil the reasons of vulnerabilities being introduced 
in dependency trees, as well as possible solutions.}

Through our empirical study, we conclude some findings as follows. For example, \textbf{1)} We statistically prove that vulnerabilities widely exist in the dependencies of NPM packages (over one-quarter of library versions from 20\% of libraries), {even in the latest versions (16\% of libraries).} Besides, vulnerabilities from direct dependencies are widely neglected (over 30\% affected library versions). \textbf{2)} Known vulnerabilities are causing a larger impact 
over time, with {more affected packages and more vulnerabilities in dependency trees}. Most vulnerabilities (93\%) are introduced into dependency trees before they are discovered, and most fixing versions (87\%) of them are released before they get published. Based on these timely releases, {most of the vulnerable dependencies can be removed} (90\%) along with time (in average it takes a year), but there are still 40\% of vulnerabilities unable to get thoroughly excluded. {\textbf{3)} Considerable user projects contain unavoidable vulnerabilities even though we have exhausted all possible dependency trees.}
More findings can be found in \Cref{sec:empirical_study}.
Additionally, since the severe situation of vulnerability propagation is complicated and requires efforts from different roles to mitigate, we also conclude actionable solutions from different stakeholders in the supply chain of third-party libraries based on our findings in~\Cref{sec:discussion}.

In summary, we make the main contributions as follows.

\begin{itemize}[leftmargin=*]
    \item We design and construct a complete and precise \textsf{DVGraph} for the whole NPM ecosystem {by leveraging a robust dependency constraint parser}. The construction and maintenance pipelines take 20 person-months. 
    {\item We propose a novel algorithm (\textsf{DTResolver}) based on DVGraph to statically and {precisely} resolve the dependency trees for any installation time with high accuracy (over 90\%), which is validated by around 100k representative packages.} 
    \item We conduct the first large-scale empirical study based on {over 50 million resolved dependency trees (calculated on an 8-core machine for one month)} to peek into the vulnerability propagation and the evolution of vulnerability propagation over time and provide useful findings.
    \item We provide an in-depth discussion, including lessons learned and actionable solutions, which provide useful insights to improve the security of the whole NPM ecosystem for different stakeholders, such as the proposed remediation (\textsf{DTReme}) that excludes more vulnerabilities than the official tool \textsf{npm audit fix}.
    \item We have made the relevant analytic data publicly available\footnote{\url{https://sites.google.com/view/npm-vulnerability-study/}} to facilitate the relevant research on the NPM ecosystem.
\end{itemize}

\Cref{fig:approach} demonstrates the overview of our work, including the dependency-vulnerability knowledge graph construction (\Cref{sec:graph}), dependency tree resolution, vulnerable path identification, and their validations (\Cref{sec:dep_resolution}), a large-scale empirical study (\Cref{sec:empirical_study}) and discussion on lessons learned and solutions, as well as possible research directions (\Cref{sec:discussion}).

\section{Motivating Example and Background}\label{sec:background}

\subsection {\textbf{Motivating Example}}\label{subsec:motivating}
\begin{figure}
\centering
\includegraphics[width=0.45\textwidth]{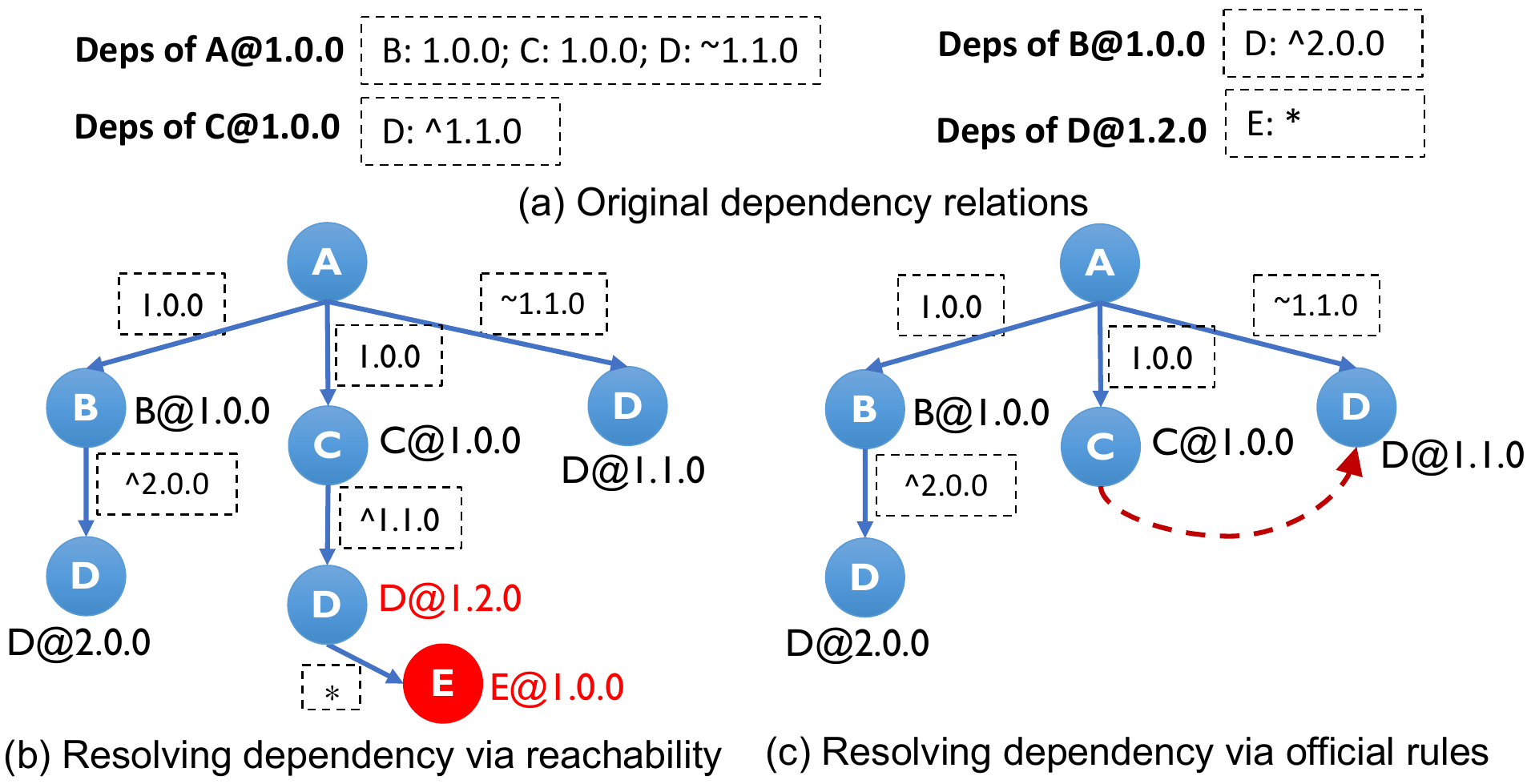}
\footnotesize
    \begin{enumerate}
      \item {D has four versions: 1.0.0, 1.1.0, 1.2.0, 2.0.0}
      \item {According to \textsf{node-semver}, $^\sim${$1.1.0$} represents ``<1.2.0 and >=1.1.0'', \^{$1.1.0$} represents ``<2.0.0 and >=1.1.0'', ``*'' represents any version.}
    \end{enumerate}
\caption{An example of NPM dependency resolution}
\label{fig:npmexample}
\end{figure}

{Here, we present an example to illustrate why it is unreliable to conduct vulnerability propagation analysis via existing reachability analysis.}
\Cref{fig:npmexample} presents an example, where each package is represented in the format of \textsf{library@version}.
\Cref{fig:npmexample} (a)
presents the original dependency relations of packages in the format of \textsf{library:constraint}.  
\Cref{fig:npmexample} (b) and \Cref{fig:npmexample} (c) present the resolved dependency of A@1.0.0 via reachability reasoning (i.e., dependency reach in~\cite{zimmermann2019small}) and NPM official rules, respectively. 
Specifically, when resolving the dependency of C@1.0.0, reachability analysis selects D@1.2.0 because 1.2.0 is the highest satisfying version of dependency constraint \^{$1.1.0$} from C@1.0.0 to D, and package E@1.0.0 is also selected. 
However, this may lead to inaccurate results. 
NPM follows its own principle to resolve dependencies during installation~\cite{npmalgo}. For example, NPM takes \textsf{real installation context} into account when resolving \textsf{dependency trees} (e.g., allowing existing versions to be reused), while resolving transitive dependency via reachability fails to involve such rules. As presented in \Cref{fig:npmexample} (c), since existing D@1.1.0 satisfies
\^{$1.1.0$}, it will directly reuse D@1.1.0 instead of resolving a new one. Thus, D@1.2.0 and E@1.0.0 are wrongly resolved by reachability analysis in \Cref{fig:npmexample} (b).

\subsection{Background}

We briefly introduce several concepts related to dependency management and vulnerabilities in the NPM ecosystem.

\smallskip
\noindent \textbf{NPM Package Metadata.}
NPM~\cite{npmjs.org} is the official package manager for Node.js and provides a public NPM registry 
to maintain information of all published libraries (i.e., \textsf{NPM package metadata}~\cite{npmpackagemetadata}). 
Such metadata describes all information of a given library and its released versions, and the dependency relation of each version, which is useful for indexing libraries and resolving dependencies when installing a library.

\smallskip
\noindent \textbf{Dependency Constraints.}
{Dependencies of package $p$} are specified as a list of key-value pairs in metadata, {where \textsf{key} represents a library $lib_a$ and \textsf{value} represents the {allowed version range (i.e., constraint) that $p$ should follow when selecting the version of library $lib_a$ during installation.}} 
%
{There are different types of constraints in NPM~\cite{npm-pkg-arg}, such as \textsf{Version} and \textsf{Range}, \textsf{Tag},  \textsf{URLs} (i.e., {Git} urls and {Remote} links), and local paths (\textsf{Directory and File}).}

\smallskip
\noindent \textbf{Semantic Versioning.}
It has been proposed as a solution of version control when maintaining package dependencies.
It defines version numbers presented in the format of \textsf{Major.Minor.Patch}. {When a new version is released, }\textsf{Major} increases when incompatible API changes have been taken out, \textsf{Minor} raises when functionality changes are still backward compatible, and \textsf{Patch} grows when backward compatible bug fixes have been made.


\smallskip
\noindent \textbf{Vulnerabilities in NPM Packages.}
Vulnerabilities in NPM packages are included in the CVE reports. Each CVE is published with detailed information about vulnerabilities and their references. {Specifically, the affecting libraries and the corresponding versions are always described in free text descriptions, which raises more efforts to retrieve and map CVEs with exact affecting library versions.}

\section{DVGraph Construction}\label{sec:graph} 


To support large scale dependency-based vulnerability analysis with high accuracy and {efficiency}, we design and implement a set of infrastructures to construct and maintain a complete and precise dependency-vulnerability graph (\textsf{DVGraph}). 

\begin{figure}
	\centering
	\includegraphics[width=0.45\textwidth]{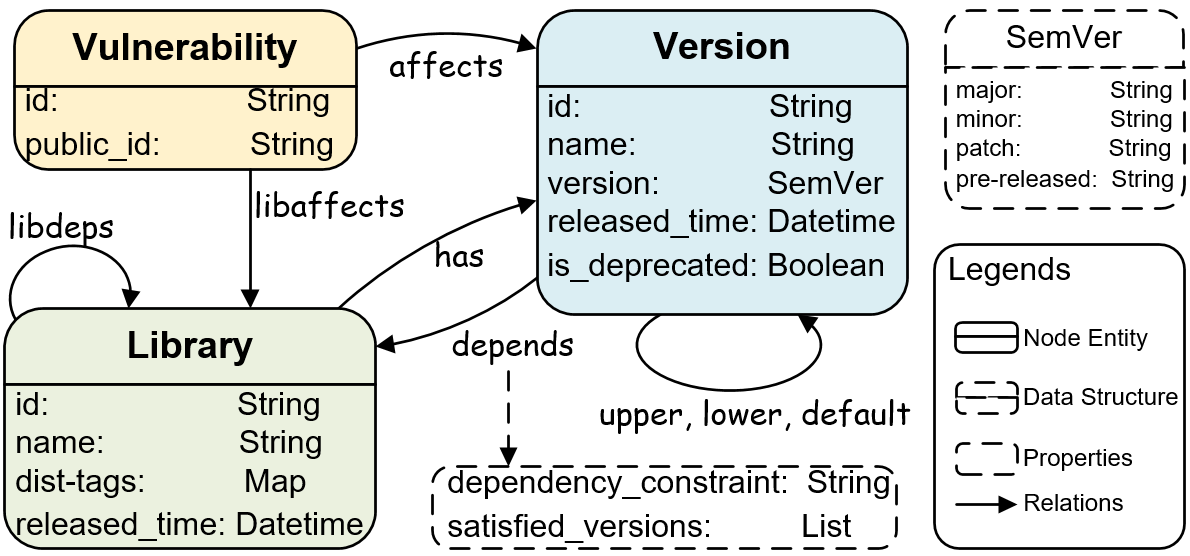} 
	\caption{Schema of NPM dependency-vulnerability graph}
	\label{fig:schema}  
\end{figure}

\smallskip
\noindent \textbf{DVGraph Definition and Schema.}
The NPM $DVGraph$ is designed as a directed knowledge graph with labeled vertices and directed edges: $G = (N, E)$, where $N$ represents
all node entities with different types in $G$, i.e., Library (\textit{Lib}), Version (\textit{Ver}), and Vulnerability (\textit{Vul}).
$E$ represents the relations between nodes, 8 types in total, 
{including inner-library relations (i.e., \textit{has}, \textit{upper}, and \textit{lower}), cross-library relations (i.e., \textit{depends}, \textit{default}, and \textit{libdeps}), and vulnerability-related relations (i.e., \textit{affects} and \textit{libaffects}).}
\Cref{fig:schema} shows the detailed {schema} of each type of nodes and relations.
{Specifically, we present relations in the format of ``<{Node}>--<\textit{relation}>$\rightarrow$<{Node}>'' as follows:}
\ding{182}~\textsf{Inner-library relations}: ``$\rm Lib_1$--\textit{has}$\rightarrow\rm {Ver}_1$'' denotes $\rm {Lib}_1$ has a released version $\rm {Ver}_1$, ``$\rm {Ver}_1$--\textit{upper/lower}$\rightarrow\rm{Ver}_2$'' denotes the next semantically upper/lower version of $\rm {Ver}_1$ is $\rm {Ver}_2$.
\ding{183}~\textsf{Cross-library relations}:  ``$\rm {Ver}_1$--\textit{depends}$\rightarrow\rm {Lib}_2$'' denotes $\rm {Ver}_1$ directly depends on $\rm {Lib}_2$.
    Specifically, {for each $depends$, we denote ``$\rm {Ver}_1$--\textit{default}$\rightarrow\rm{Ver}_2$'', where $\rm {Ver}_2$ is the latest version of $\rm {Lib}_2$ that satisfies the $dependency\_constraint$ of $depends$ 
    }, because NPM takes the latest satisfied version when resolving dependency constraint to specific versions by default. Moreover, ``$\rm {Lib}_1$--\textit{libdeps}$\rightarrow\rm {Lib}_2$'' denotes that there exists at least one version of $\rm {Lib}_1$ depending on $\rm {Lib}_2$.
\ding{184}~\textsf{Vulnerability-related relations}: Since vulnerabilities usually 
    exist in multiple versions, ``$\rm Vul_1$--\textit{affects}$\rightarrow\rm Ver_1$'' and ``$\rm Vul_1$--\textit{libaffects}$\rightarrow\rm Lib_1$'' denote that $\rm Vul_1$ exists in $\rm Ver_1$ and exists in at least one version of $\rm Lib_1$, respectively.

\begin{figure}
	\centering
	\includegraphics[width=0.48\textwidth]{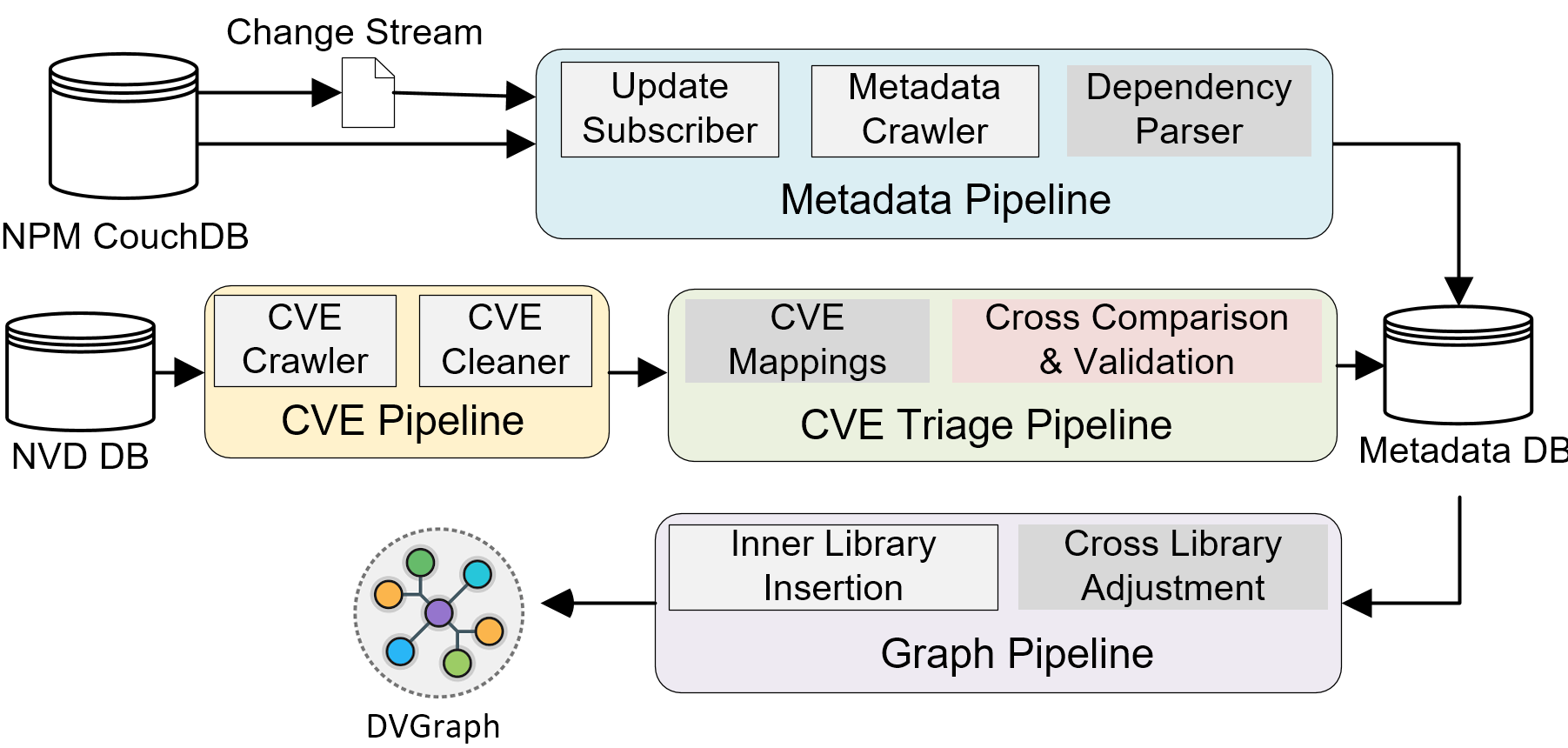} 
	\caption{Automated data processing framework}
	\label{fig:dataflow}  
\end{figure}

\smallskip
\noindent \textbf{DVGraph Construction and Maintenance.}
The DVGraph is constructed as a knowledge graph in Neo4j~\cite{neo4j}.
{NPM resolves dependency constraints with the highest satisfying version by default, which claims a high demand for real-time updates of dependency data.} Therefore, we conduct an automated data processing framework for long-term maintenance as shown in ~\Cref{fig:dataflow}, including:
(1) \textsf{Metadata Pipeline}, subscribes, daily collects, cleans, and processes new coming NPM package metadata, and preserves them in our metadata database.
(2) \textsf{CVE Pipeline}, collects CVE feeds~\cite{cvefeeds} from the NVD database. Since some information in CVE feeds are usually in plain text, a CVE cleaner is designed to filter the languages and identify the affected libraries, as well as affecting version ranges, and save them as the initial results.
(3) \textsf{CVE Triage Pipeline}, is a semi-automated pipeline. It helps experienced security analysts process, label and confirm the newly crawled CVE data with corresponding affected libraries and versions.
(4) \textsf{Graph Pipeline}, parses the new coming metadata and mapped CVE data, calculates the operations (i.e., adding, altering, and deleting nodes and edges) to be done on DVGraph, and finally executes them. 

Specifically, there are two challenges during \textsf{DVGraph} update.

\begin{itemize}
    \item {\textbf{Dependency Parser}: The diversity of NPM dependency constraints makes it complex to resolve proper versions. 
    Wrongly handled dependency relations can cause deviation when reasoning 
    transitive dependencies,} 
    and none of the existing work has taken all major types into account. To this end, we propose and develop \textbf{a robust dependency constraint parser} based on \textit{node-semver}~\cite{node-semver}. It handles not only semver version ranges but also version tags, Git URLs, and remote links~\cite{npm-pkg-arg}.
    \item {\textbf{CVE Mappings}: Even though the CVE pipeline crawls and processes CVE data and automatically recognizes affected libraries and version ranges, there still could be mislabeling since they are usually given in free-text descriptions~\cite{Guo2021PMA,guo2021key}. Therefore, we implement the {CVE Triage Pipeline} as shown in \Cref{fig:dataflow} and have devoted four experienced analysts {to check the CVE mappings of affected library and version ranges,} some existing famous vulnerability databases~\cite{snykcves}~\cite{sourceclearcves} are also involved as references.} {After confirmation, all affected versions can be sorted out by the constraint parser.}
\end{itemize}


\begin{tcolorbox}[size=title,opacityfill=0.1,breakable]
Remarks: It takes 3 months for 3 co-authors and experienced software engineers to implement and test, and another 4 security analysts to conduct daily validation of CVE mappings.
\end{tcolorbox}

\begin{table}
\footnotesize
\renewcommand{\arraystretch}{1}
\caption{Graph statistics}
\label{tbl:graphstatistics}
\centering
\scalebox{1}{\begin{tabular}{lc|lc}
\hline
\textbf{Elements} & \textbf{\#Instances} & \textbf{Elements} & \textbf{\#Instances} \\ \hline
{\textit{Lib}} & 1,147,558 & {\textit{has}} & 10,939,334  \\
{\textit{Ver}} & 10,939,334 & {\textit{upper}} & 9,804,406 \\
{\textit{Vul}} & 815 & {\textit{lower}} & 9,804,406 \\ \hline	
{\textit{depends}} & 62,232,906 & {\textit{affects}} & 23,217    \\
{\textit{default}} & 61,940,009 & {\textit{libaffects}} & 830\\
{\textit{libdeps}} & 4,216,742 & Graph size & 15.15GB \\
\hline
\end{tabular}}
\end{table}

\noindent \textbf{DVGraph Statistics.}
{To carry out large-scale studies on 
NPM vulnerability propagation and evolution,} we take a snapshot of \textsf{DVGraph} 
{by the end of 2020 (Most of CVEs before 2020 are finalized)}
to conduct further analysis. Table~\ref{tbl:graphstatistics} shows the basic statistics of the snapshot. 1,147,558 libraries and 10,939,334 versions have been captured in the \textsf{DVGraph}, among which 62,232,906 direct dependencies ($depends$) are captured in DVGraph. Besides, 815 CVEs have been included in DVGraph, generating
830 \textit{libaffects} and 23,217 \textit{affects} from these CVEs to 624 libraries and 14,651 versions, respectively. Overall, the storage size of the DVGraph snapshot is over 15GB. 

\smallskip
\noindent \textbf{DVGraph and CVE Mapping Validations.}
{To roughly validate the coverage of \textsf{DVGraph}, we take a snapshot of the metadata database to compare with it}. We find that DVGraph covers 100\% of libraries and 99.96\% of versions of the metadata database (the rest are unpublished~\cite{npmunpublish}). 
Besides, only 0.36\% of direct dependencies are not captured in \textsf{DVGraph} because the corresponding dependency libraries are missing in the NPM registry. Furthermore, we also find 0.47\% of $depends$ cannot be resolved to any satisfying version due to no satisfying version or invalid dependency constraints. 


\begin{tcolorbox}[size=title,opacityfill=0.1,breakable]
Remarks: As for the validation of CVE mappings, besides 4 security analysts have validated all new-coming CVEs every day, {the co-authors} still manually {check} the mappings between all CVEs and the affected library versions after we conduct the snapshot. In particular, since the publish time of CVEs differs from several sources~\cite{nvd,cve,snykcves}, we take the earliest ones.
\end{tcolorbox}

\section{Dependency Tree Resolution and Vulnerable Path Identification}\label{sec:dep_resolution}
To facilitate large-scale studies, we propose a novel methodology on dependency tree resolution to statically resolve the dependency trees of packages based on the complete and well-maintained {DVGraph}. It not only resolves dependency trees precisely but also preserves the high {efficiency} of static analysis, {which enables most of SCA~\cite{sca} scanning tasks to be carried out without real installation, e.g., license violation detection, untrustworthy dependency detection, etc.} Therefore, identifying vulnerabilities and corresponding vulnerable paths that vulnerable packages propagate to affect the root package can also be accurately and efficiently captured.

\begin{figure}
	\centering
	\begin{subfigure}[t]{0.24\textwidth}
		\centering
		\includegraphics[width=0.83\textwidth]{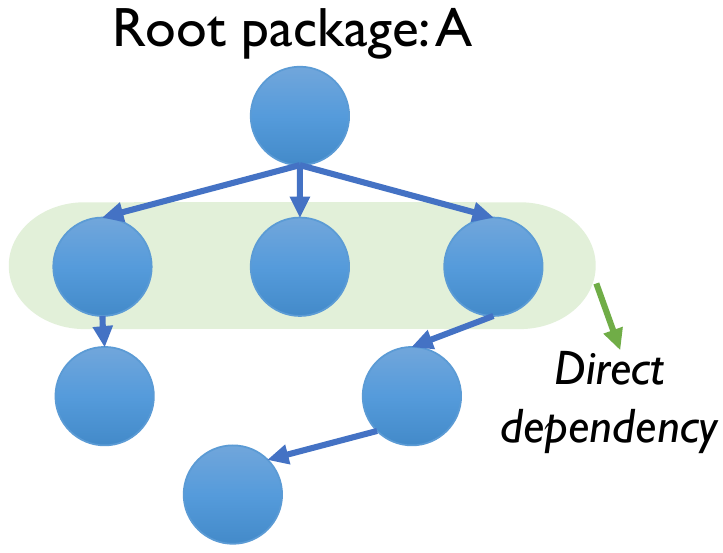}
	\caption{Dependency tree of package \textit{A}}
		\label{subfig:deptree}
	\end{subfigure}%
	\begin{subfigure}[t]{0.27\textwidth}
		\centering
		\includegraphics[width=0.96\textwidth]{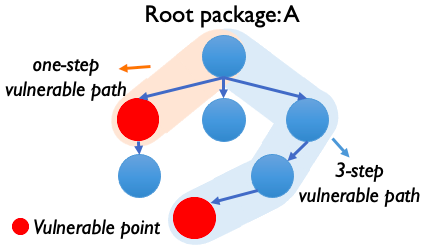}
	\caption{Vulnerable Paths}
		\label{subfig:vulpath}
	\end{subfigure}%
	\caption{Examples of dependency tree and vulnerable paths (each node represents a package with an exact version)}
	\label{fig:treepath}
\end{figure}

\subsection{Dependency Tree Definition}
Third-party libraries are usually installed along with the installation of user projects as dependencies, during which NPM follows its own dependency resolution rules and resolves dependency constraints into specific library versions recursively. These resolved library versions and dependency relations among them form a directed dependency graph.

{Precisely, we denote the graph as \textsf{\textbf{dependency tree}}, $DT_{root} = (Ver, Dep)$, $root$ denotes the root package {(could be user project)} that is to be installed, and $Ver$ represents the set of library versions that are resolved during the real installation, and $Dep$ is the set of dependency relations among $Ver$. To distinguish with $depends$ in \textsf{DVGraph}, we further define $v_i$ $\overset{dep}{\rightarrow}$ $v_j$ as the specific dependency relation between library version $v_i$ and $v_j$ in $DT$. {{\Cref{subfig:deptree} shows the dependency tree $DT_A$ of package \textit{A} starting from package \textit{A} (the root package)} and connecting all resolved library versions with dependency relations.} The $Vers$ {that are directly depended on by} the root package A are denoted as \textbf{direct dependencies}, and the others that are not directly depended on by A are denoted as \textbf{indirect (transitive) dependencies}, as presented in \Cref{subfig:deptree}.}

\subsection{Dependency Tree Resolution}\label{subsec:dependency_resolution}
Lots of work~\cite{wittern2016look,kula2017impact,decan2016topology,lertwittayatrai2017extracting} has been carried out to investigate transitive dependencies in the NPM ecosystem. However, none of them have taken into consideration the platform-specific dependency resolution rules~\cite{resolution_rules}, 
{which could result in inaccurate dependencies being resolved} (illustrated by the example shown in \Cref{fig:npmexample}).
{To fill this gap, we aim to {\textbf{statically}} resolve dependency trees that are consistent with {what NPM dynamically resolves and installs during real installation,} so that we can identify vulnerabilities and vulnerable paths in dependency trees precisely {and efficiently} without real installation.}

Besides the reachability analysis, there exist other tools~\cite{npmaudit} scanning vulnerabilities by examining dependencies after real installation. 
However, such dynamic approaches always take a much longer time than static approaches and are not efficient enough for large-scale analysis.
{Besides, during installation, NPM manipulates dependencies based on physical trees {(connecting $Vers$ based on physical location of installation)}, which makes the inner dependency relations in logical trees (connecting $Vers$ {based on} dependency relations) implicit. 
Therefore, to improve the accuracy and meanwhile reserve the efficiency, we propose a DVGraph-based dependency resolution algorithm (\textsf{DTResolver}) to {statically} calculate the implicit logical {dependency} trees {without installation}.}

Specifically, as presented in \Cref{algo:depresolve}, we simulate the folder allocation process~\cite{resolution_rules} during real installation, 
{and recursively resolve dependencies for each selected library versions (lines {5$\sim$27}). }
$Q$ denotes a queue to control the recursion process, and $Dir$ represents a virtual and empty directory (as well as an empty physical tree, line 1).
For each visited library version in $Q$, we iterate its dependencies alphabetically {(line 8)}. 
For each dependency relation, the latest satisfied version will be selected when resolving a certain dependency {(line 15)}. However, such resolving is also influenced by the physical tree. If there is an installed version $v_i$ for the required library that is on a higher position in the physical tree, the installed one will be reused instead of resolving a new version {(lines 11$\sim$13)}. {Otherwise, a new latest satisfied version will be resolved and installed {(lines 15$\sim$27)}. Note that, we denote $install\_path$ to present install position in the physical tree, and when $path\_1$ $\sqsubseteq$ $path\_2$, it means $path\_2$ is inside $path\_1$. In other words, $path\_1$ is on higher position of $path\_2$ {(lines 11, 20)}.}

Meanwhile, we also maintain the individual dependency relations between packages during the simulation process {(lines 12, 25)}. 
Therefore, we can further recover the logical tree with the physical tree structure and those dependency relations. 
Besides, we also include lots of individual version selection rules (e.g., priorities on deprecated versions~\cite{npmdeprecatedpriority} and latest tags~\cite{npmlatest}) that NPM follows in our algorithm to the best of our knowledge.

Additionally, we extend our algorithm to \textbf{time dimension} by adding filters on release time when selecting satisfying versions {(comments on {line 15}), and this empowers the calculation of dependency trees at any previous time from release time.
Therefore, more time-based analyses could be carried out.} 

\SetKwInput{KwInput}{Input}
\SetKwInput{KwInput}{Input}
\SetKw{Let}{let}
\SetKw{Continue}{continue}
\SetKw{Break}{break}
\SetKw{Create}{CREATE}
\begin{algorithm2e}[t]
    \footnotesize
	\setcounter{AlgoLine}{0}
	\caption{{Dependency Tree Resolution}}
	\label{algo:depresolve}
	\DontPrintSemicolon
	\SetCommentSty{mycommfont}
	{
	    \KwIn{$G$: DVGraph, $r$: given root package, \tcp*{$t$: given time}}
    	\KwOut{$DT_r$: Resolved dependency tree of r}
        $Dir$ $\gets$ new InstallDirectory()\;
        $root\_path \gets \emptyset$, $Q$ $\gets \emptyset$, $Deps$ $\gets$ $\emptyset$\;
        $Dir$.install($r$, $root\_path$)\;
        $Q$.push($r$)\;
        \tcp*[h]{1. Traverse all resolved dependency nodes by BFS, and simulate real installation to create folders for packages}\;
        \While {$Q$ $\neq$ $\emptyset$}{
            $lv$ $\gets$ $Q$.pop()\;
            $deps$ $\gets$ $\{e \in G: {e_{src} = lv \wedge e.type = depends}\}$\;
            \ForEach{$depend$ $\in$ $deps$}{
                $vers$ $\gets$ $depend.satisfied\_versions$\;
                deplib $\gets depend_{dst}$\;
                \eIf {$\exists$ $v_i$. $v_i$ $\in$ Dir $\cap$ vers $\wedge$ $v_i.dir\_path$ $\sqsubseteq$ $lv.dir\_path$}{
                    $r$ $\gets$ ($\Create$ $lv \overset{dep}{\rightarrow} v_i$)\;
                    $Deps$.push($r$)\;
                }{
                    $selected \gets v_i. v_i \in vers \wedge (\forall v_j. v_j \in vers \wedge i \neq j \wedge v_i > v_j)$\tcp*{$\wedge v_i.released\_time < t$}\;
                    \eIf{Dir $\cap$ vers = $\emptyset$}{
                        $install\_path \gets root\_path$\;
                    }
                    {
                        \ForEach{subpath $\sqsubseteq$ lv.dir\_path}{
                            \If {$\neg \exists n. n \in subpath \wedge (deplib-has\rightarrow n$)}{
                                $install\_path \gets subpath$\;
                                $\Break$ \;
                            }
                        }
                    }
                    $Dir$.install($selected$, $install\_path$)\;
                    $r$ $\gets$ ($\Create$ $lv \overset{dep}{\rightarrow} selected$)\;
                    $Deps$.push($r$)\;
                    $Q$.push($selected$)\;
                }
            }
        }
        \tcp*[h]{2. Recover a dependency tree from install directory and CREATED Deps relations}\;
        $Ver_{r}$ $\gets$ \{$lv: lv \in Dir$\}\;
        $Dep_{r}$ $\gets$ $Deps$\;
        $DT_{root}$ $\gets$ <$Ver_{r}$, $Dep_{r}$>\;
        \Return $DT_{r}$\;
    }

\end{algorithm2e}

\subsection{\textbf{Vulnerable Path Identification}}
Since all known vulnerabilities (CVEs) and their affecting libraries and versions are already well mapped in \textsf{DVGraph}, 
the resolved dependency trees can further provide the capability to extract dependency paths. 
{Dependency paths connect $Vers$ in $DT$ in series via $Deps$ among them, each path $P$ is a subset of $DT$ with orders, and can be denoted as $P = (PN, PE)$, $P$ $\sqsubseteq$ $DT$, where $PN$ is an ordered list of $Vers$ that can be connected via $Deps$ in the given order, and $PE$ is the ordered list of $Deps$ that connect $Vers$, they satisfy Equation \ref{equation:path_node} and \ref{equation:path_edge}, respectively.} 
\begin{multline}
\forall~i. ~~i \in [0, k-1) \wedge v_i \in PN \vdash \\
\exists~d. (d \in Dep \wedge d_{src} = v_i \wedge d_{dst} = v_{i+1}) \label{equation:path_node}
\end{multline}
\begin{flalign}
\forall~e_i. (e_i \in PE) \vdash
e_{i\_src} = v_i \wedge e_{i\_dst} = v_{i + 1}
\label{equation:path_edge}
\end{flalign}
\begin{multline}
P|^{src}_{dst} = \{P_i \sqsubseteq DT: 
PN_{i\_0} = v_{src} \wedge PN_{i\_k} = v_{dst})\} 
\label{equation:last_node}
\end{multline}
{where $k$ denotes the length of $PN$, and $e_i$ denotes the dependency relation between $v_i$ and $v_{i+1}$. Specifically, there could be multiple paths to one target node in a dependency tree. Considering the cross dependencies, we denote the set of dependency paths from library version $v_{src}$ to library version $v_{dst}$ as $P|^{src}_{dst}$, as defined in Equation \ref{equation:last_node}. Specifically, we denote the set of dependency paths from root package to library version $v_{dst}$ as $P|_{dst}$.}

{Based on the resolved dependency trees and paths for the installed packages, the algorithm also facilitates the identification of vulnerable packages in dependencies (i.e., \textbf{vulnerable point}) and the paths that these vulnerable packages propagate on to affect the root package (i.e., \textbf{vulnerable path}), and we also denote them as Equation \ref{equation:affection_point} and \ref{equation:vulnerable_path}, respectively. $AP_A$ denotes the set of $Vers$ in $DT_A$ that are affected by known vulnerabilities, $VP$ denotes the set of $P_i$ whose last node $P_{i\_k-1}$ is affection point, $Vul$ denotes all vulnerabilities we have maintained in DVGraph. Moreover, we define \textbf{K-step vulnerable path} $VP_{S=k}$ as vulnerable paths that contain $k$ dependency relations, as defined in Equation \ref{equation:k_step}.} 
\begin{multline}
AP_A = \{v \in Ver_A:\\
\exists n_{vul}. (n_{vul} \in Vul \wedge (n_{vul}-affects \xrightarrow{} v))\} 
\label{equation:affection_point}
\end{multline}
\begin{flalign}
VP_A = \{P|_v \sqsubseteq DT_A: (P|_v)_{k-1} \in AP_A\} 
\label{equation:vulnerable_path}
\end{flalign}
\begin{flalign}
VP_{S=k} = \{P|_v \in VP_A: |(PE|_v)| = k\}
\label{equation:k_step}
\end{flalign}

{Examples of vulnerable points and paths are given in \Cref{subfig:vulpath}.} Thus, we implement a vulnerable path extractor by reverse Depth First Search (DFS) to exhaustively find dependency relations from vulnerable points to the root node in a dependency tree.

\subsection{Validation of Dependency Tree Resolution}\label{subsec:validation}
We validate the \textsf{DTResolver} by comparing the dependency trees resolved by \textsf{DTResolver} with the real installed ones. Moreover, we take \textsf{npm-remote-ls}~\cite{npmremotels} as a baseline method when comparing, which is a widely-used public API to get dependency trees without real installation in practice{, and it exactly follows the dependency reach to derive dependency trees.}

\smallskip
\noindent \textbf{Data Selection}
Our validation is based on the data collected by two criteria: {(1) \textsf{Popularity}, for each popularity metrics (i.e., most stars, most forks, most downloaded in the past, past 3 years, and last year), we select the top 2,000 libraries respectively.
(2) \textsf{Centrality}, for each centrality metric (i.e., most in and out degree), we also select the top 2,000 libraries and top 20K versions. respectively.} For libraries, we take the highest patch version for each minor version. Finally, 103,609 versions from 15,673 libraries are sorted out. 

\smallskip
\noindent \textbf{Experiment.}
Based on the collected data, we first collect all installation dependency trees (\textsf{Install Tree}) for each version {from real installation ({\textsf{npm-install}~\cite{npminstall} and} \textsf{npm-ls}~\cite{npmls})},
82,415 of these versions are successfully collected after excluding those with installation errors.
We also collect the dependency trees (\textsf{Remote Tree}) from \textsf{npm-remote-ls}. Moreover, to compare with real installation results,
we update the graph after all \textsf{Install Trees} are well collected, so that all packages in the \textsf{Install Trees} are updated into the graph. Based on it, we further compute the dependency trees (\textsf{Graph Tree}) for all versions with their corresponding installation times. 

\smallskip
\noindent \textbf{Evaluation of \textsf{DTResolver}.} According to the result, \textbf{90.58\%} of \textsf{Graph Trees} exactly match the \textsf{Install Trees} after ignoring incalculable cases, e.g., having bundled dependencies~\cite{npmbundled} and containing dependencies with no released time.
While only 53.33\% of \textsf{Remote Trees} exactly match the \textsf{Install Trees}, which is because {\textsf{npm-remote-ls} have missed some official resolution rules (e.g., priority selection on not deprecated versions).} {Besides, we further identified two major reasons for mismatched dependency trees: 1) Dependencies are deduplicated~\cite{npmdedupe} in the output of \textsf{npm ls}, which omits some packages and dependency relations to simplify the tree view. 2) Dependencies may not be fully installed due to environmental issues (e.g., some packages may not be installed when the required OS support is missing). Besides, missing library versions (i.e., not in the NPM registry or crawling failure) also cause some missing packages in the dependency trees.}

\smallskip
\noindent \textbf{Evaluation of Vulnerability Detection and Vulnerable Path Identification.}
{Besides the evaluation of \textsf{DTResolver}, we also extend to compare the detected vulnerabilities and vulnerable paths. Since the \textsf{Install Tree} retrieved from real installation may be incomplete (e.g., some packages in dependencies are not installed due to environment issues), we evaluate the accuracy of vulnerability detection by the recall of the identified vulnerabilities and vulnerable paths in \textsf{Graph Tree} and \textsf{Remote Tree}. We find that both DTResolver (98.1\%) and \textsf{npm-remote-ls} (97.7\%) have similarly high coverage on detecting vulnerable components but vary on identifying vulnerable paths (92.60\% vs. 78.31\%). This is probably because most dependency constraints are resolved to the highest satisfied version, and dependency reach also follows this rule, therefore, most vulnerable packages can still be identified. However, resolving dependencies via dependency reach 
neglects the NPM specific resolution rules, which compromises the accuracy on identifying dependency paths.}

{The results not only prove the quality of the \textsf{DVGraph} and the accuracy of \textsf{DTResolver}, but also the accuracy of vulnerability detection and vulnerable path identification.}
We take more evaluation details and case analysis on our website (\url{https://sites.google.com/view/npm-vulnerability-study/}).

\section{Large-Scale Empirical Study}\label{sec:empirical_study}
\newcounter{finding}
\newcounter{RQ}
The dependency reachability reasoning adapted in existing work makes it difficult to carry out more fine-grained analysis on dependencies, such as deducing the dynamic changes of vulnerability in dependency trees, due to the neglecting of inner dependency relations (i.e., structure) in dependency trees as discussed in ~\Cref{subsec:dependency_resolution}. However, such analysis is vital to unveil the reasons and characteristics of vulnerabilities being introduced as dependencies to support precise remediation for dependency trees and even solutions to mitigate the entire NPM ecosystem. 
Therefore, our proposed \textsf{DTResolver} is vital, and based on this, we further carry out our study on vulnerability propagation and evolution in the context of dependency trees from these two research questions:

\begin{itemize}
    \item \textbf{RQ1}: (\textsf{Vulnerability Propagation via Dependency Trees}) How do vulnerabilities affect the NPM ecosystem? 
    How do vulnerabilities propagate to affect root packages via dependency tree?
    \item \textbf{RQ2}: (\textsf{Vulnerability Propagation Evolution in Dependency Trees}) How do vulnerability propagation evolves in dependency trees? How do dependency tree changes influence the evolution of vulnerability propagation? 
\end{itemize}

\subsection{RQ1: Vulnerability Propagation via Dependency Trees}\label{subsec:impact}
\label{subsec:rq2}

The goal of this section is to investigate how does NPM dependency resolution influence the vulnerability propagation via dependency trees from two aspects:
1) the propagation of vulnerability via dependency trees and 
2) the influence on vulnerability propagation brought by NPM dependency resolution.

\subsubsection{\textbf{
{How many packages are affected by existing known vulnerabilities in the NPM ecosystem?}
}}\label{subsubsec:how_many_libs}
Existing studies~\cite{decan2018impact, zimmermann2019small} have unveiled that vulnerabilities in third-party libraries can widely affect the NPM ecosystem via dependencies, while their neglecting on NPM specific dependency resolution rules may lead to inaccurate dependencies (\textit{cf.}  Figure~\ref{fig:npmexample}), resulting in biases in conclusions.
{Therefore, we re-evaluate the 
vulnerability impact 
by computing dependency trees for all packages in the NPM ecosystem and analyzing vulnerability propagation for each of them.}



As presented in Table~\ref{tbl:graphstatistics}, we have captured 815 known vulnerabilities in \textsf{DVGraph}, which exist in {14,651 versions} from 624 libraries. The amount of these library versions (\textsf{directly affected}) are relatively small, comparing to the mass of the NPM ecosystem. However, based on the dependency trees we resolved, we find that \textsf{an astonishing portion (i.e., 24.78\%, 2,711,222) of versions, from 19.96\% (229,037/1,147,558) libraries, are transitively affected by 416 CVEs, which are introduced from versions of 294 vulnerable libraries}. 

Besides, since users are always recommended to take the latest version of libraries to get rid of vulnerability, we further analyze the vulnerability {propagation} in the latest versions of all libraries (1,147,558), and we find that the latest versions of 185,598 libraries (16.17\%) are still transitively affected. 
This finding reveals that latest versions of third-party libraries are also under potential risk of being affected by known vulnerabilities via dependencies.

{Moreover, we further notice a bad practice that the latest version of 35.03\% (103/294) of vulnerable libraries that have dependent packages are vulnerable.} Since NPM usually resolves dependency constraints as the highest satisfying versions, these vulnerable latest versions have a much higher chance of being depended on by other packages than old versions, leading to much higher possibility of distributing indirect vulnerability {propagation}.


\begin{tcolorbox}[size=title,opacityfill=0.1,breakable]
\textbf{Finding-1}: \ding{172} It is statistically proved that vulnerabilities are widely existing in dependencies of NPM packages (one-quarter versions of 19.96\% libraries across the ecosystem).
\ding{173} Latest versions of third-party libraries (16.17\%) are still under potential risks of being affected by vulnerabilities via dependencies.
\ding{174} A considerable portion of vulnerable libs (over 100) that are used by others, still have vulnerable latest versions.
\end{tcolorbox}

\subsubsection{\textbf{How do vulnerabilities propagate to affect root packages via dependency tree?}}\label{subsubsec:how_to_propa}

{Based on the dependency trees (over 10 million) we resolved, we also extract the vulnerable points and corresponding vulnerable paths to investigate how do vulnerabilities propagate to affect root packages via dependency trees.}

For vulnerable points, we notice that there is clear centrality that some influential vulnerabilities transitively affect a significant portion of library versions in the NPM ecosystem. Particularly, we find 25 CVEs have affected over 10k libraries or 100k versions (1\% of the entire ecosystem), which might be utilized to threaten the NPM ecosystem.
The top 10 of them are presented in Table~\ref{tbl:top10CVE}. 

\begin{table}[!h]
\footnotesize
\centering
\caption{Top 10 CVEs that affect most versions}
\label{tbl:top10CVE}
\scalebox{0.95}{
\begin{tabular}{lccc}
\hline
{\textbf{Public ID}}& {\textbf{Source Lib.}} & {\textbf{\#Affected Ver.}} & {\textbf{\#Affected Lib.}}\\
\hline
{CVE-2019-10747}& {\textbf{set-value}}& {948,208}& {73,947}\\
{CVE-2019-10744}& {\textbf{lodash}} & {867,148} & {79,459}\\
{CVE-2018-16487}& {lodash}& {819,360}& {77,433} \\
{CVE-2018-3721}&{lodash} &{790,100} & {75,817} \\
{CVE-2018-3728} & {hoek} &{741,754} & {62,227}\\
{CVE-2019-1010266} & {lodash} &{712,971} & {70,956}\\
{CVE-2018-1000620} & {cryptiles} &{601,414} & {52,334}\\
{CVE-2018-20834} & {tar} &{592,691} & {48,356}\\
{CVE-2017-16137} & {debug} &{509,455} & {38,626}\\
{CVE-2016-10540} & {minimatch} &{388,126} &{41,423}\\
\hline
\end{tabular}}
\end{table}

{To have a more intuitive view of vulnerability propagation, we extract vulnerable paths from these vulnerable points to corresponding root packages. Note that we ignore library versions (0.5\%) which have over 1k vulnerable paths, since it takes too much time to compute all paths exhaustively.}
Finally, we identified \textbf{88,192,572 vulnerable paths}.
On average, each vulnerable version has 3.97 vulnerable points in its dependency tree, which generate 32.53 vulnerable paths. Besides, nearly 90\% of vulnerable paths are longer than 3 steps. This means that each vulnerable point in dependency trees may have multiple complex paths to affect the root packages, and these results prove that remediation on single dependency relation or even single vulnerable path may not be enough to exclude corresponding vulnerabilities completely.

Besides, it is surprising that there are still 33.33\% vulnerable versions (903,569) having \textbf{one-step vulnerable paths} {(refer to \Cref{subfig:vulpath})}, and 12.04\% of them (326,404) only have such paths. 
Since direct dependencies are visible to developers and maintainers as configured in \textsf{package.json}~\cite{npmpackagejson},
{one-step vulnerable path} 
should be easily identified and handled if developers and maintainers are sensitive to vulnerabilities in dependencies. 
Moreover, we further notice that one-step vulnerable paths exist in dependency trees of 33.42\% (62,022/185,598) of vulnerable latest versions,
which means even for the latest versions, developers and maintainers have not paid enough attention to security in dependencies. 
{These findings indicate a universal lack of attention on vulnerabilities from dependencies, even for vulnerabilities from direct dependencies.}

{As for multi-step vulnerable paths, the vulnerable paths of 49.57\% of vulnerable versions (1,344,020) propagate and affect root package via only one direct dependency. 
Vulnerable paths of 78.94\% of them (2,140,239) go through no more than 3 direct dependencies. 
This indicates that {most vulnerable paths are centralized to propagate and affect root packages via limited direct dependencies}, and it proves that controlling direct dependencies precisely may be an effective solution to cut off most vulnerable paths.}


\begin{tcolorbox}[size=title,opacityfill=0.1,breakable]
\textbf{Finding-2}: \ding{172} There are centrality that some influential known CVEs widely exist in the dependency trees of a significant portion of packages.
\ding{173} Packages are usually affected by multiple vulnerable points, and each vulnerable point affects root packages via multiple vulnerable paths (averagely, one vulnerable points introduce 8 vulnerable paths).
\ding{174} Vulnerabilities still widely exist in direct dependencies of affected library versions (over 30\%), even for the latest versions.
\ding{175} There is also centrality on vulnerable paths that most of the vulnerable paths go through limited direct dependencies, which could be utilized to cut off vulnerable paths.
\end{tcolorbox}

\subsection{RQ2: Vulnerability Propagation Evolution in Dependency Trees}\label{sebsec:rq3}
\begin{figure}
\centering
\includegraphics[width=0.45\textwidth]{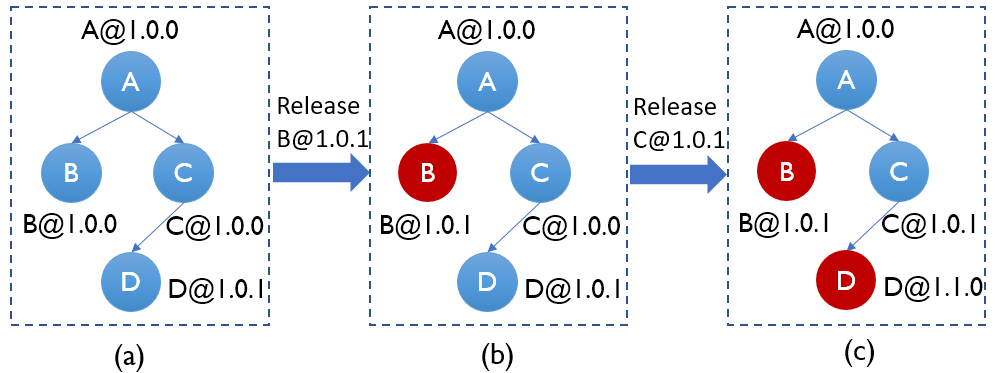}
\caption{An example of vulnerability {propagation} evolution via dependency tree changes (DTCs)}
\label{fig:vulintroexample}
\end{figure}

{Since dependency trees installed by default could change along with the release of new version of any library in the tree,} 
it is highly possible that the status of root packages being affected by vulnerability via dependencies also changes over time. An example of dependency tree changes (\textsf{DTCs}) that introduces vulnerability is depicted in \Cref{fig:vulintroexample}. A@1.0.0 has experienced two DTCs when B and C release new versions, two vulnerable points (B@1.0.1 and D@1.1.0) are introduced into the dependency tree of A@1.0.0.
{Note that the dependency trees we analyze are the ones to be installed by default, instead of the outdated dependencies in runtime environment.}

To investigate the evolution of vulnerability {propagation}, therefore, find out reasons for the widespread of vulnerability propagation and further derive actionable solutions, we resolve dependency trees and analyze corresponding vulnerable paths over time to investigate: 
1) the overall evolution of known vulnerability propagation in dependency trees; 
2) the lifecycle of vulnerabilities in dependency trees;
3) a possible solution to mitigate vulnerability affection in dependency trees based on our findings.

Due to the exponential increase of dependency trees over time, we conduct our analysis on the validation dataset in \Cref{subsec:validation}.
Overall, 53,541 versions are selected after excluding versions with unusual dependency trees by Quartile Variation~\cite{bonett2006confidence} and those that have dependencies with unknown release times, and we further calculate their dependency trees from release time to the latest. Finally, we obtain  
10,906,781 dependency trees in total.

\begin{figure}
    \centering
\begin{subfigure}[t]{0.235\textwidth}
	\includegraphics[width=0.98\textwidth]{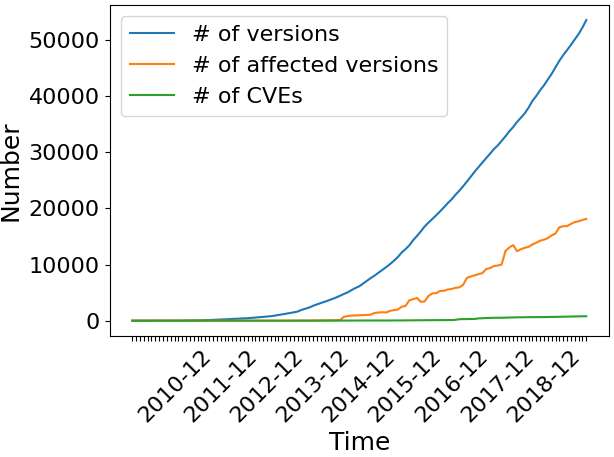}
	\caption{Evolution of library versions and CVEs}
	\label{subfig:snapshot}
\end{subfigure}
\hfill
\begin{subfigure}[t]{0.235\textwidth}
	\includegraphics[width=0.98\textwidth]{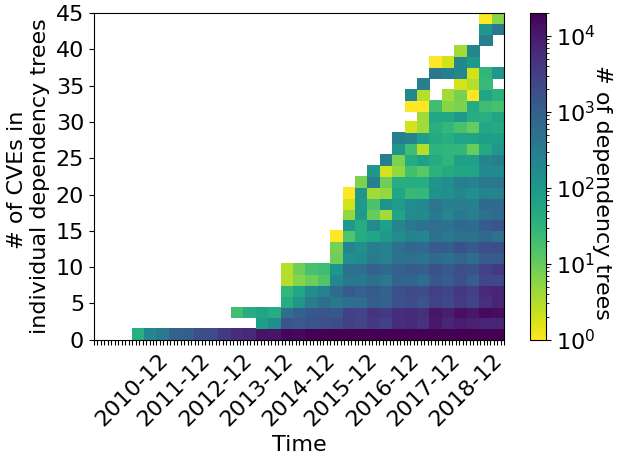}
	\caption{Evolution of CVE density in dependency trees}
	\label{subfig:density}
\end{subfigure}
\caption{Evolution of known CVE propagation}
\end{figure}

\subsubsection{\textbf{How does known vulnerability {propagation} evolve over time?}}\label{subsubsec:known}

To investigate the historical vulnerability {propagation} in dependency trees, we first take snapshots of dependency trees 
at the end of each month
for all the 53,541 versions.
Note that here we only measure the propagation of known vulnerabilities from their publish time.

As presented in \Cref{subfig:snapshot}, library versions and known CVEs grow rapidly, which is consistent with existing work~\cite{decan2018impact,zimmermann2019small}. Besides, we further investigate known vulnerabilities in dependency trees to analyze the breadth of known vulnerability propagation. 
Specifically, we find 19.27\% of these versions (10,320) were affected by known vulnerabilities at release time, while 33.86\% of versions (18,127) are affected by known vulnerabilities at the latest time. This indicates that more versions are getting affected by known vulnerabilities from dependency trees over time.

Besides, we analyze the changes of the number of CVEs in individual dependency trees over time. 
As shown in \Cref{subfig:density}, the number of CVEs in dependency trees increases rapidly over time.
Specifically, we identify 69.76\% of these ever affected versions (14,356) have more known vulnerabilities in dependency trees at the time of snapshot comparing to release time. In contrast, only 7.4\% of them (1,524) are just the opposite. Our findings indicate that, along with the discovery of known vulnerabilities and DTCs over time, {the impact of known vulnerabilities is getting larger that known vulnerabilities are not only affecting more library versions (breadth), but also affecting each version via multiple vulnerable points (depth). 


\begin{tcolorbox}[size=title,opacityfill=0.1,breakable]
\textbf{Finding-3}: Known vulnerabilities are causing a larger impact across the NPM ecosystem over time. They are not only affecting more library versions but also affecting them with more vulnerable points in dependency trees.
\end{tcolorbox}

\subsubsection{\textbf{How long do vulnerabilities live in dependency trees?}}\label{subsub:lifecycle} Vulnerabilities in dependency trees can cause enormous impact not only after published. To investigate the evolution of vulnerability {propagation} (i.e., lifecycle of vulnerabilities) in dependency trees, 
{we analyze how vulnerability changes in dependency trees.}

Therefore, we take the time when the root package was affected by vulnerabilities as the initial time, and further define the living time of each CVE in dependency trees as the time interval between \textbf{CVE introduction} (the first DTC that brings this CVE {to the dependency tree}) and \textbf{CVE elimination} (the last DTC that removes this CVE {from the dependency tree})}.

We have identified 213 CVEs from the 10 million dependency trees, and quantified 243,448 individual CVE introductions related to these CVEs. Besides, to look into the living time of these CVEs in dependency trees, we have further quantified corresponding CVE eliminations for these introduced CVEs. As a result, 60.05\% of them (146,192) are removed from the corresponding dependency trees afterward, with an average living time of 371 days, while the rest of them (39.95\%) still remain in the latest dependency trees.
Specially, we find that 87.69\% of CVE eliminations (128,190) happen before these CVEs are published, while only 7.4\% of CVE introductions happen after CVE publish time.

These findings reveal that {most CVEs (around 93\%) are introduced to dependency trees before they are discovered, and half of the introduced CVEs (60\%) in dependency trees are removed via DTCs. Besides, 88\% of such CVE eliminations happen before CVE publication since most CVEs only get published after the maintainers have fixed them. However, the living time of these removed CVEs is still longer than a year, and there is still 40\% of CVEs in dependency trees not removed.} This is probably because it indeed requires a quite long time to handle CVEs (e.g., identifying security bugs, fixing these bugs, publishing CVEs). However, the remaining CVEs prove that there still exist CVEs that can not be removed by DTCs automatically, and lacks mature mechanisms to warn users of the CVEs in dependencies of their projects efficiently. 


\begin{tcolorbox}[size=title,opacityfill=0.1,breakable]
\textbf{Finding-4}: \ding{172} Most of the CVEs (93\%) have already been introduced to dependency trees before they were discovered, and the fixed versions of these CVEs (87\%) were also mostly released before CVE publish.
\ding{173} Only 60\% of CVEs in dependency trees are removed automatically by DTCs, and even so, it still takes over one year for each CVE to get removed.
\end{tcolorbox}

\subsubsection{\textbf{Why are there still a considerable portion of CVEs not removed?}}\label{subsec:notremoved}
Actually, most of the CVEs can be removed by DTCs over time, and the remained CVEs are the ones we should try to mitigate.
To investigate the possible reasons of the remaining 40\% of CVEs {and further conclude applicable solutions,} {we extract all vulnerable paths introduced by these CVEs from dependency trees and classify
them into \textbf{Not Removed Paths (NRP)} (i.e., those remained in latest dependency trees) and \textbf{Removed Paths (RP)} (i.e., those removed by DTCs).} 
Therefore, 318,652 \textsf{NRPs} and 1,669,258 \textsf{RPs} have been identified.
{According to our previous findings, the latest versions being vulnerable are probably the common features of remained vulnerable points, and}
 the root cause of CVE introduction and {elimination} is the change of dependency trees, which requires two preconditions: 1) nodes in the dependency tree have new versions released; 2) the newly released version satisfies the corresponding dependency constraint. {{Therefore, we think there are probably the two main reasons} that may break preconditions and block {CVE eliminations}: \textsf{Outdated Maintenance} (i.e., no clean version released) and \textsf{Unsuitable Dependency Constraint} (i.e., the released clean version does not satisfy the dependency constraint)}.

We also select two typical cases to analyze how they block CVE {eliminations}:
1) \textsf{Vulnerable Latest Version}, the highest satisfying version of vulnerable point is vulnerable; (2) \textsf{Fixed-Ver. D.C.}, the dependency constraints are fixed versions instead of ranges. We measure both of them in \textsf{NRPs} and \textsf{RPs}.
The result shows that 1) {Compared to \textsf{RPs},} \textsf{NRPs} {contain more} Fixed-Ver. D.C. (53.10\% vs. 25.67\%), and {have more} 
Fixed-Ver. D.C. per path (0.87 vs. 0.43).
2) Vulnerable points in 61.54\% of \textsf{NRPs} are Vulnerable Latest Versions, while only 16.12\% of \textsf{RPs} are in such cases.
{These findings prove that bad practices such as \textsf{Outdated Maintenance} (provider) and \textsf{Unsuitable Dependency Constraint} (consumer) are possibly the reasons that postpone or even block {the automatic removal of CVEs by DTCs over time}, and more effective and actionable instructions and solutions should be further derived against them.}


\begin{tcolorbox}[size=title,opacityfill=0.1,breakable]
\textbf{Finding-5}: \textsf{Outdated Maintenance} (provider) and \textsf{Unsuitable Dependency Constraint} (consumer) are the main reasons that {hinder} the automated vulnerability removal in dependency trees over time. More countermeasures and solutions should be carried out to avoid, monitor, or even fix these bad practices.
\end{tcolorbox}

\subsubsection{\textbf{Example of remediation by avoiding vulnerability introduction} (\textsf{DTReme})}
\textsf{Outdated Maintenance} and \textsf{Unsuitable Dependency Constraint} are the main reasons that hinder CVEs from being automatically removed, which requires 
all stakeholders' efforts to exclude. However, we can remediate vulnerabilities
{from another direction by preventing CVE introductions.} Therefore, 
we further propose and implement an application to provide a novel and more precise remediation (\textsf{DTReme}) for dependency trees, based on the \textsf{DTResolver} we presented in~\Cref{subsec:dependency_resolution}.

Theoretically, vulnerable paths could be introduced or removed by DTCs when new versions of libraries in the middle of paths are released. Therefore, we could use forward checking~\cite{haralick1980increasing} and backtracking~\cite{dechter1998backtracking} to explore all possible solutions for single dependency path and avoid resolving vulnerable versions, therefore, avoiding the introduction of vulnerabilities. However, since dependency paths are not independent and could be influenced by other existing nodes in dependencies, fixing single vulnerable path may not be able to remove the vulnerability thoroughly (i.e., these remained vulnerable points in~\Cref{subsec:notremoved}). 
Therefore, we combine NPM dependency resolution and strategies of forward vulnerability checking and backtracking to resolve clean dependency trees, thus, provides remediation for entire dependency trees. 

Notably, we add 1) forward vulnerability checking, when resolving versions for new coming dependencies (line 13 and 17 in Algorithm~\ref{algo:depresolve}), only resolve clean versions for every dependency relation; 2) backward installed package tracking, once no clean version could be resolved, roll back to the resolution for parent node and find alternative versions to avoid cases like no clean version. Therefore, we can traverse all possible solutions exhausitively and find possible clean dependency trees, and a new \textsf{package-lock.json} file {can be generated for the entire dependency tree as the remediation solution.} Note that the integrity issue of lock file is handled by \textsf{ssri}~\cite{ssri}.

To prove the effectiveness, we evaluate \textsf{DTReme} with popular JavaScript repositories from Github by comparing the remediation result with \textsf{npm audit fix}, {the official dependency auditing tool}.
We first collect top 1K most stared repositories from Github~\cite{github}, after excluding unsuitable projects (115 have no \textsf{package.json}, 239 use yarn, 27 have no dependencies, 159 have dependencies that are not published in NPM registry), we obtain 460 projects as experiment objects to compare with \textsf{npm audit fix}.
Next, we collect 3 types of dependencies, default dependencies (\textsf{DefDep}), dependencies after audit fix (\textsf{AuditDep}), and dependencies after remediation (\textsf{RemeDep}), as well as vulnerable points in these dependency trees to compare remediation effects. The results are presented in \Cref{tbl:remediation}. 

{Overall, our \textsf{DTReme} handles more vulnerabilities than \textsf{npm audit fix}.}
Among the 262 projects that have vulnerabilities in their dependencies, 
the performance of our DTReme is better than \textsf{npm audit fix} in 77 projects (i.e., the deep gray cell), while only 30 projects (i.e., the light gray cell) are opposite. However, these 30 cases are because that sometimes \textsf{npm audit fix} remediates vulnerabilities by violating direct dependency constraints~\cite{auditbug}, and our remediation follows user-defined dependency constraint strictly. 
Besides, among the 155 projects that \textsf{DTReme} and \textsf{npm audit fix} have the same performance, \textsf{DTReme} reduces more vulnerable paths introduced by these vulnerable points in 16 projects.

\begin{tcolorbox}[size=title,opacityfill=0.1,breakable]
\textbf{Finding-6}: These results prove that \textsf{DTReme} has better performance on remediation than \textsf{npm audit fix}. Back tracing the vulnerable paths to the status before vulnerabilities are introduced is an effective way to exclude more vulnerabilities in dependency trees. 
Besides, these results also prove that there are noticeable vulnerabilities unavoidable even though we have exhausted all possible dependency paths, and mitigating such vulnerabilities requires all stockholders to be responsible for their parts and working together.
\end{tcolorbox}

\begin{table}
\small
\caption{Comparison of remediation effects between \textsf{npm audit fix} and our remediation}
\label{tbl:remediation}
\centering
\begin{tabular}{lc}
\toprule
\textbf{\# of vulnerable points in Dependency Trees}             & \textbf{\# of projects} \\ \hline
DefDep = 0                                                                  & 198                                 \\
DefDep = AuditDep = RemeDep \textgreater 0                                  & 86 (15)                                  \\
DefDep \textgreater AuditDep = RemeDep                                      & 69 (1)                                  \\
DefDep \textgreater{}= AuditDep \textgreater RemeDep                        & \cellcolor{graytwo}{77}                                  \\ 
DefDep \textgreater{}= RemeDep \textgreater AuditDep                       & \cellcolor{grayone}{30}                                  \\ 
\bottomrule
\end{tabular}
\end{table}

\section{Discussion}\label{sec:discussion}
\renewcommand{\thefootnote}{$\star$}

\subsection{Lessons Learned by Our Study}\label{subsec:implications}

We discuss the actionable solutions from different stakeholders to mitigate the severe situation.

\smallskip
\noindent{\textbf{For Package Providers}}.
%
Outdated maintenance is one of the major reasons for CVEs remaining in dependency trees shown in \Cref{subsec:notremoved}.
Thus, we conclude some tips:
\ding{172} releasing patch versions soon when vulnerabilities are found{, especially for those major versions that are not latest by still widely used}; 
\ding{173} deprecating or unpublishing the vulnerable versions from the NPM registry; 
\ding{174} being more responsible to maintain at least one satisfying clean version for most commonly used dependency constraints, 
especially when moving to the next major versions; 
\ding{175} frequently checking the dependencies of their own packages with additional tools (e.g., third-party auditors), and remediating vulnerable dependencies in time, in case they propagate to transitively affect downstream users.

\smallskip
\noindent \textbf{For Package Consumers}. 
%
We recommend managing dependency trees with a compromised strategy, using dependency lock with periodically updating dependency trees to include new features and vulnerability fixes {that are still compatible}. Such a strategy could trade off the conflict with limited risks of containing known CVEs and reduced compatibility issues. 
Besides, there is a noticeable lack of attention on vulnerabilities in dependencies. A significant portion of the transitively affected packages contain one-step vulnerable paths as shown in \Cref{subsubsec:how_to_propa}. 
We strongly call on the attention of users on vulnerabilities in dependencies, especially in direct dependencies, and more analysis tools (i.e., third-party auditors) should also be applied during software development and maintenance.

\smallskip
\noindent \textbf{For Third-party Auditors.} 
Most existing software component analysis (SCA) tools are heavy (i.e., require real installation or lock file). Lighter static tools (e.g., \textsf{DTReme}) can be further included (e.g., in IDEs) and help developers examine their dependencies with much higher frequency. 
Besides, there are more directions based on our findings that could improve the security for users' projects.
\ding{172} \textsf{More fine-grained remediation.}
Apart from patching on vulnerable codes, currently, version-based remediations are manipulating either vulnerable points (\textsf{npm audit fix}) or direct dependencies only (e.g., snyk's remediation~\cite{snykfix}), while there are still lots of CVEs in dependency trees unremediated.
As presented by \textsf{DTReme}, simply excluding vulnerable dependencies from path level is efficient to remediate much more vulnerabilities than \textsf{npm audit fix}, and more recommendations, e.g., even replacing libraries with similar functionalities, could be further investigated.
\ding{173} \textsf{More accurate reachability analysis.} Package-level detection is not accurate and could introduce false positives~\cite{8530065}. 
Although it is a difficult task~\cite{moller2020detecting,kristensen2019reasonably,stein2019static}, reachability analysis based on call graph~\cite{nielsen2021modular} can precisely filter out if these vulnerable codes are really called.

\subsection{Limitations and Threats to Validity}
First, the vulnerabilities in dependencies may never actually affect root packages since these vulnerable functions may never be reached. This can only be further tackled by analyzing vulnerable function call paths based on dependency trees and call graphs. We leave this as our future work.
Second, the mapping of CVEs and library versions is labeled manually, which may cause data mislabeling, and the co-authors have cross-validated the data with existing CVEs to mitigate such threats.
{Third, we can not distinguish installations that contain missing dependencies, which could make the ground truth inaccurate, we only take packages in dependencies that are successfully installed as ground truth in validation.}
Fourth, we ignore versions with over 1k vulnerable paths when analyzing vulnerability propagation due to excessively high computation costs. Overall, such versions only account for 2.01\%, which can only cause limited bias to our results.

\balance
\section{Related work}

\noindent{\bf Vulnerability Analysis via Dependency.}
Lauinger et al.~\cite{lauinger2018thou} checked 133k websites and found 37\% websites use at least one JavaScript library with a known vulnerability.
Pfretzschner et al.~\cite{10.1145/3098954.3120928} discussed four typical dependency-based attacks.
Ohm et al~\cite{ohm2020towards} investigated the security attacks via malicious packages from supply chain.
Prana et al~\cite{prana2021out} observed the vulnerabilities from dependencies of selected projects in Java, Python and Ruby by Veracode~\cite{veracode}.
Alfade et al~\cite{alfadel2020threat} measured the threats of vulnerabilities by their lifecycle. 
Zerouali et al.~\cite{zerouali2019impact} reported that the presence of outdated NPM packages increases the risk of potential vulnerabilities. 
Gkortzis et al~\cite{gkortzis2021software} found a strong correlation between a higher number of dependencies and vulnerabilities.
Javanjafari et al~\cite{javanjafari2021dependency} investigated the dependency smells that could cause negative impact in JavaScript projects.
Decan et al.~\cite{decan2018impact} conducted an empirical study by leveraging the direct dependencies of JavaScript libraries. Their impact analysis is conducted only with {direct dependents which are upstream traced with only one step}, and did not consider the dependencies as integrity to analyze the vulnerability impact via transitive dependencies. 
{Zerouali et al.~\cite{zerouali2021impact} investigated vulnerability impact via transitive dependencies, but they reasoned such impact via dependency reach, and they are more focusing on comparison between packages and projects on general properties, i.e., dependency level.} 
Zimmermann et al.~\cite{zimmermann2019small} found the individual packages could impact large parts of the NPM ecosystem, {and they analyzed from the perspective of maintainer accounts that could be used to inject malicious code. However,~their dependencies are also derived from dependency reachability reasoning.}
Most existing work only conducted dependency-based vulnerability impact analysis on the reachability of library dependencies or limited transitive dependency steps, while our study is conducted based on dependency trees with high accuracy at a large scale.


Some work focused on improving security based on dependency.
Cox et al.~\cite{cox2015measuring} proposed a metric-based method to decide if the dependencies should be updated. 
Van et al.~\cite{van2019server} proposed {NodeSentry} to identify the vulnerable libraries by using rules for secure integration of JavaScript libraries. 

In summary, these work focused on mitigating vulnerabilities based on dependencies that are already installed, while our approach focuses on identifying vulnerabilities and vulnerable path before installation. Besides, we can also include recommendations on possible remediation for identified vulnerabilities via \textsf{DTReme}.

\noindent{\bf Ecosystem Analysis.}
In most cases, researchers analyzed the dependency relations in various languages to understand the ecosystem status and dependency evolution.
Wittern et al.~\cite{wittern2016look} investigated the NPM ecosystem from several aspects (e.g., library dependency, download metrics). 
Decan et al.~\cite{decan2016topology,7884604_2017,decan2019empirical} conducted comparison studies of different ecosystems, and they~\cite{decan2019package} also recommend recommend semantic versions by the wisdom of the crowd. 
Similarly, Kikas et al.~\cite{kikas2017structure} analyzed the dependency network and evolution of three ecosystems (i.e., NPM, Ruby, and Rust).
Besides, some existing empirical studies further analyzed different aspects of the ecosystem from various entry points. 
~\cite{lertwittayatrai2017extracting} used topological data analysis to investigate the NPM ecosystem. 
\cite{kula2017impact,abdalkareem2017developers, chen2021helping, chowdhury2021untriviality} investigated the prevalence and impact of trivial packages. 
\cite{decan2018evolution, zerouali2018empirical, zerouali2019formal, stringer2020technical, chinthanet2021lags} investigate the technical lags of adopting updates in several ecosystems.
\cite{zerouali2019diversity} compares different metrics on popularity, and \cite{dey2018software} predicts the popularity change based on dependency supply chain.
\cite{abate2020dependency} compares the dependency resolution of different package managers.
\cite{cogo2021empirical} focus on packages with the same day release.
\cite{qiu2021empirical} investigates the problem of license violation.
\cite{chatzidimitriou2019npm} inspects the packages that are usually co-occur dependencies.
\cite{chinthanet2019lag} investigated the slow patching process within the NPM ecosystem.

Compared with our study, most of the existing work on ecosystem analysis only focused on a limited group of study subjects and more focused on general statistics of entire ecosystems without precisely considering the accurate dependency reachability. Instead, we have conducted a large-scale study on a full set of packages in the NPM ecosystem, and have analyzed on both ecosystem and package (version) levels with accurate dependency trees considered.

\section{Conclusion}
In this paper, we carry out a large-scale empirical study on the vulnerability propagation and propagation evolution by leveraging a complete and precise \textsf{DVGraph}, and a novel algorithm that we firstly propose to statically and precisely resolves accurate dependency trees \textsf{DVResolver} {at any time} for each package. Our study unveils many useful findings on the NPM ecosystem. Based on it, we propose \textsf{DVReme} as an example to mitigate vulnerability impact, and we also highlight some implications to shed light on the severe security threats in NPM ecosystem and further invoke actionable solutions for different stakeholders to mitigate such security risks.

\begin{acks}
This research was partially supported by the National Natural Science Foundation of China (Grant No. 62102284),
the National Research Foundation, Singapore under its the AI Singapore Programme (AISG2-RP-2020-019), the National Research Foundation, Prime Ministers Office, Singapore under its National Cybersecurity R\&D Program (Award No. NRF2018NCR-NCR005-0001), NRF Investigatorship NRFI06-2020-0022-0001, the National Research Foundation through its National Satellite of Excellence in Trustworthy Software Systems (NSOE-TSS) project under the National Cybersecurity R\&D (NCR) Grant award no. NRF2018NCR-NSOE003-0001, the Ministry of Education, Singapore under its Academic Research Fund Tier 3 (MOET32020-0004). Any opinions, findings and conclusions or recommendations expressed in this material are those of the author(s) and do not reflect the views of the Ministry of Education, Singapore.
\end{acks}
\clearpage
\balance
\bibliographystyle{ACM-Reference-Format}
\bibliography{icse}

\end{document}